\documentclass[pdflatex,sn-mathphys-num]{sn-jnl}
\usepackage{graphicx}%
\usepackage{multirow}%
\usepackage{amsmath,amssymb,amsfonts}%
\usepackage{amsthm}%
\usepackage{mathrsfs}%
\usepackage[title]{appendix}%
\usepackage{xcolor}%
\usepackage{textcomp}%
\usepackage{manyfoot}%
\usepackage{booktabs}%
\usepackage{algorithm}%
\usepackage{algorithmicx}%
\usepackage{subfigure}
\usepackage{multicol}
\usepackage{algpseudocode}%
\usepackage{listings}%
\usepackage{geometry} 
\theoremstyle{thmstyleone}%
\theoremstyle{thmstyletwo}%

\theoremstyle{thmstylethree}%

\begin{document}

\title[Article Title]{Evaluating Supervised Learning Approaches for Quantification of Quantum Entanglement}


\author[1]{\fnm{Shruti} \sur{Aggarwal}}\email{shruti\_2k19@dtu.ac.in}

\author*[1]{\fnm{Trasha} \sur{Gupta}}\email{trashagupta@dtu.ac.in}
\equalcont{These authors contributed equally to this work.}

\author[2]{\fnm{R. K.} \sur{Agrawal}}\email{rkajnu@gmail.com}
\equalcont{These authors contributed equally to this work.}

\author[1]{\fnm{S.} \sur{Indu}}\email{s.indu@dce.ac.in}
\equalcont{These authors contributed equally to this work.}

\affil[1]{\orgname{Delhi Technological University}, \orgaddress{\state{Delhi}, \country{India}}}

\affil[2]{\orgname{Jawaharlal Nehru University}, \orgaddress{\state{Delhi}, \country{India}}}


\abstract{Quantum entanglement is a key resource in quantum computing and quantum information processing tasks. However, its quantification remains a major challenge since it cannot be directly extracted from physical observables. To address this issue, we study a few machine-learning based models to estimate the amount of entanglement in two-qubit as well as three-qubit systems. We use measurement outcomes as the input features and entanglement measures as the training labels.  Our models predict entanglement without requiring the full state information. This demonstrates the potential of machine learning as an efficient and powerful tool for characterizing quantum entanglement.}




\maketitle

\section{Introduction}\label{sec1}

  Entanglement is a quantum mechanical feature that has no classical analogue. Quantum entanglement serves as a fundamental resource for a wide range of quantum protocols and applications, such as quantum superdense coding \cite{PhysRevLett.69.2881}, quantum teleportation \cite{PhysRevLett.70.1895}, and quantum cryptography \cite{PhysRevLett.67.661}. Consequently, various measures have been proposed to quantify the entanglement present in quantum systems \cite{RevModPhys.81.865, Ma2011_GME, aggarwal2025classification}. In the case of bipartite systems, several entanglement measures have been introduced, including entanglement of formation \cite{woottersprl98}, concurrence \cite{hillprl97}, and negativity \cite{vidalpra2002}. \\
The quantification of entanglement in multipartite systems poses a formidable challenge, arising from the richer and more intricate mathematical structures involved compared to bipartite systems \cite{kimpra2010,krammerprl2009}. For bipartite pure states, quantum entanglement can be fully characterized using the Schmidt decomposition \cite{schmidt1907theorie}. This provides a local unitary canonical form, which separates the state’s local parameters from its nonlocal ones. The Schmidt coefficients encode all nonlocal properties and, in principle, allow any entanglement measure to be expressed in terms of them \cite{bennetprl96}. In contrast, for multipartite pure states, no general local unitary canonical form exists, making the quantification of multipartite entanglement a more challenging and intriguing problem \cite{tarrachprl2000}.\\
Entanglement is not a physical observable and cannot be directly measured in experiments. Conventional approaches of quantification require full quantum state tomography, which becomes exponentially costly with the number of qubits, and many entanglement measures involve convex-roof optimizations that are computationally intractable for mixed states \cite{Koutny2023}. These limitations make the experimental quantification of entanglement a formidable challenge. Machine learning offers an efficient alternative by learning the mapping between experimentally accessible data, such as expectation values of observables, and entanglement measures \cite{Wang2025}. Once trained, such models can provide fast and reliable estimates without full state reconstruction, enabling scalable characterization of quantum correlations \cite{feng2024quantifying, mahdian2025entanglement, huang2025direct}.\\
In this work, we develop a machine learning (ML) framework to quantify bipartite and multipartite entanglement in quantum systems. We construct datasets of two-qubit and three-qubit quantum states. For two qubits, concurrence is used as the entanglement measure, while for three qubits, the GME concurrence quantifies genuine tripartite entanglement. We use five supervised learning models: Support Vector Machine for Regression (SVM-R)\cite{vapnik1996support}, Decision Trees (DT-R)\cite{breiman2017classification}, Artificial Neural Network (ANN-R)\cite{specht1991general}, Generative Adaptive Model (GAM)\cite{friedman2001greedy}) and LS Boost-based Ensemble Model (LS-ENS)\cite{friedman2001greedy}). Correlation-based features derived from local observables and Svetlichny operators are employed as inputs to different ML models. The models are trained and optimized to predict entanglement measures from these features. Our approach provides an efficient and scalable method for entanglement quantification in quantum information processing.\\
This paper is organized as follows: In Section-2, some basic definitions of quantum states and a few entanglement measures are stated. In Section 3 and 4, related work and methodology are stated. Section 5 contains the experimental setup and Section 6 contains the results obtained and discussion. Finally we conclude in Section 7.

\section{A Note on Quantum Entanglement}
Let $\mathcal{H} \equiv \mathcal{H}_A \otimes \mathcal{H}_B$ be composite Hilbert space where $A$ and $B$ are finite-dimensional subsystems.
A bipartite quantum state $\rho_{AB}$ is defined as
    $\rho_{AB} = \sum_i p_i |\psi_i\rangle \langle\psi_i| \;\;\text{with}\; \sum_i p_i = 1 \; \text{and} \; p_i \geq 0$.  When it is known that the quantum system is in the state $|\psi_k\rangle$, then is said to be in pure state. In this case the density operator is simply $\rho = |\psi_k\rangle\langle\psi_k|$.
 $\rho_{AB}$ is a product state if there exist states $\rho_{A}$ for subsystem $A$ and $\rho_{B}$ for subsystem $B$ such that
$\rho_{AB} = \rho_{A} \otimes \rho_{B}$;
The state $\rho_{AB}$ is separable if it can be written as a convex combination of product states, i.e., $\rho_{AB} = \sum_k p_k \, \rho_A^k \otimes \rho_B^k, \;\; \sum p_k = 1 \label{sepdef}\; \text{and}\; 0 \leq p_k \leq 1 \; \forall \; k$ Otherwise $\rho_{AB}$ is an entangled state.\\
Acín et al. introduced a classification scheme for three-qubit states into distinct entanglement classes~\cite{acin2001}. These include separable states, biseparable states, and genuine entangled states which consist of the W class, and the GHZ class, where
$|GHZ\rangle = \tfrac{1}{\sqrt{2}} \left( |000\rangle + |111\rangle \right);\;\;\; |W\rangle = \tfrac{1}{\sqrt{3}} \left( |100\rangle + |010\rangle + |001\rangle \right)$. The GHZ and W classes are stochastically inequivalent, meaning that no state in one class can be converted into the other by stochastic local operations and classical communication (SLOCC).
\subsection{A Few Entanglement Measures}
We begin with a discussion of two analytic measures of quantum entanglement, namely, \textit{Concurrence} for bipartite systems and \textit{GME Concurrence} for multipartite systems.
\subsubsection{Concurrence}
Concurrence is a well-defined quantitative measure of entanglement. For the two-qubit case, an elegant formula for the concurrence was derived analytically by Wootters \cite{woottersprl98}.
Let us consider a pure state 
$|\psi_{AB}\rangle$ in the Hilbert space $\mathcal{H}$. 
The concurrence of the state $|\psi_{AB}\rangle$ is defined as $C(|\psi_{AB}\rangle) = \sqrt{2\left(1 - \mathrm{Tr}\,\rho_A^2\right)}$
where $\rho_A$ is the reduced density matrix obtained by tracing out subsystem $B$. 
This definition naturally extends to mixed states through the convex roof construction. For a mixed state, concurrence is defined as
$C(\rho_{AB}) = \min_{\{p_i, |\psi_i\rangle\}} \sum_i p_i \, C(|\psi_i\rangle)$,
where the minimization is taken over all possible pure-state decompositions defined in Section-2
For a two-qubit mixed state, concurrence is defined as $C(\rho_{AB})= max(0,\sqrt{\lambda_{1}}-\sqrt{\lambda_{2}}-\sqrt{\lambda_{3}}-\sqrt{\lambda_{4}}) \label{defconc2qubit}$ where	
    $\lambda_{i}$ are the eigenvalues of $\rho\tilde{\rho}$ arranged in descending order and $\tilde{\rho}$=$(\sigma_{y}\otimes\sigma_{y})\rho^*(\sigma_{y}\otimes\sigma_{y})$; $\sigma_{y} = - i|0\rangle\langle1| + i|1\rangle\langle0|$.
A two-qubit state $\rho_{AB}$ is separable if and only if 
$C(\rho) = 0$.\\
This construction highlights the strength of concurrence as an entanglement measure: 
it vanishes precisely for separable states, is invariant under local unitary operations, 
and provides a quantitative indicator of bipartite quantum correlations. 
However, a closed formula of concurrence is not known for higher dimensional quantum systems \cite{aggarwal2021witness}. Moreover, evaluating concurrence for mixed states remains computationally demanding 
due to the minimization over all ensemble decompositions. This motivates us to search for efficient numerical and machine learning--based approaches for the quantification of entanglement.

\subsubsection{Genuine Multipartite Entanglement (GME) Concurrence}
For an $n$-partite pure state $|\psi\rangle$, the genuine multipartite entanglement (GME) concurrence provides a quantitative measure of entanglement that captures correlations beyond bipartite partitions. 
It is defined as~\cite{Ma2011_GME}:
\begin{equation}
C_{\mathrm{GME}}(|\psi\rangle) = \min_{\gamma} \sqrt{2 \left( 1 - \mathrm{Tr}\,\rho_{\gamma}^2 \right)},
\end{equation}
where the minimization runs over all possible bipartitions $\gamma$ of the system, and $\rho_{\gamma}$ 
denotes the reduced density matrix associated with one part of the bipartition. 
This definition ensures that $C_{\mathrm{GME}}(|\psi\rangle)$ vanishes for all biseparable pure states.

By definition, $C_{\mathrm{GME}}(\rho) = 0$ if and only if $\rho$ is biseparable, and it is strictly positive for genuine multipartite entangled states. Furthermore, GME concurrence is invariant under local unitary operations and does not increase under LOCC, making it a valid entanglement monotone.

\section{Related Work}
This section conducts a review of the state-of-the-art methods used for quantifying quantum entanglement using traditional ML approaches. The researchers have investigated the capability of various ML models to characterize entanglement in the quantum system. The research work \cite{pan2024quantifying} used Artificial Neural Network (ANN) to characterize entanglement in 2-qubit and 3-qubit-based states in terms of Concurrence and GME concurrence, respectively. 
Their approach highlights the versatility and power of ANNs in capturing complex nonlinear quantum state features via partial measurement data, facilitating practical entanglement quantification.
Feng et al. \cite{feng2024quantifying} compared multiple regression models, including Random Forests and XGBoost, for predicting entanglement measures such as coherent information and relative entropy, highlighting their scalability to higher dimensions.
The research works \cite{lin2023quantifying} proposed a hybrid quantum-classical machine learning framework for entanglement quantification using Coherent Information and Relative Entropy. They extracted features like correlation data and moments from randomly sampled two qubit states. Their framework demonstrated robustness against noise. 
Huang et al. \cite{huang2025direct} developed ML-based protocols employing fully connected and LSTM networks for static and dynamic quantum states, respectively, achieving efficient and noise-resilient entanglement estimation.

\section{Methodology}\label{sec2}
This section elaborates on the ML framework used to quantify the entanglement in two and three-qubit system.

\subsection{Dataset generation}
For the supervised learning framework, we generated datasets of two-qubit and three-qubit quantum states.
\begin{itemize}
    \item{\textbf{Two-qubit states:}} The first dataset consists of two-qubit quantum states, systematically generated to ensure an even distribution of concurrence values over the range 
$[0,1]$. A total of 100,000 states were sampled, including 10,000 separable states (with zero concurrence) and 90,000 entangled states, evenly distributed across ten bins of concurrence, each of width 0.1. Within every bin, half of the states were generated as pure states, and the remaining half as mixed states, ensuring statistical balance. For each quantum state, a set of nine correlation features derived from its density matrix were computed, and the corresponding concurrence value was assigned as the target label.
    \item {\textbf{Three-qubit states:}} For the three-qubit case, a dataset of 100,000 pure states was generated and labeled using the GME concurrence as the target quantity. The data were distributed across ten bins of GME concurrence, with 10,000 separable states (GME = 0) and 90,000 entangled states in the range $(0,1]$. The entangled states comprised 50$\%$ GHZ-class and 50$\%$ W-class pure states, ensuring diverse coverage of multipartite entanglement structures. Each state was represented by eight Svetlichny-type correlation features, forming the feature vector used for training the machine learning model.
\end{itemize}
All pure three-qubit states were generated using normalized complex Gaussian vectors, a standard numerical implementation of Haar-uniform sampling on the Hilbert space. Mixed quantum states were generated using the Wishart ensemble, where a random complex Gaussian matrix $A$ is drawn and the density matrix is constructed as $\rho = \frac{A A^{\dagger}}{Tr[A A^{\dagger}]}$ .
 This method produces density matrices distributed according to the Hilbert–Schmidt measure, ensuring unbiased sampling over the state space.
\subsection{Feature Extraction}
Instead of working directly with density matrices, which contain redundant degrees of freedom, we use correlation matrices derived from expectation values of tensor products of Pauli operators.\\
For each two-qubit state $\rho_{AB}$, we compute the nine correlation features given as follows:
\[
T_{ij} = \mathrm{Tr}\!\left(\rho_{AB} \, (\sigma_i \otimes \sigma_j)\right),
\quad i,j \in \{1,2,3\},
\]
where $\sigma_{1,2,3}$ are the Pauli matrices. 

For three-qubit states, tensor products of Pauli operators across the three
subsystems are used. Expectation values of the form
$
\langle \sigma_{i} \otimes \sigma_{j} \otimes \sigma_{k} \rangle
= \mathrm{Tr}\!\left(\rho_{ABC} \, (\sigma_i \otimes \sigma_j \otimes \sigma_k)\right),
\quad i,j,k \in \{1,2,3\}$
have been computed and compressed into an 8-dimensional feature vector. 
To quantify genuine multipartite correlations, we adopt as input features the expectation values of the Svetlichny operator, which provides a device-independent criterion for detecting genuine tripartite nonlocality \cite{PhysRevD.35.3066}. The Svetlichny operator is defined as
\begin{eqnarray}
    S &=& A \otimes B \otimes C 
+ A \otimes B \otimes C' 
+ A \otimes B' \otimes C 
+ A' \otimes B \otimes C  \nonumber\\
&&- A' \otimes B' \otimes C'
- A' \otimes B' \otimes C
- A' \otimes B \otimes C'
- A \otimes B' \otimes C', \label{svetop}
\end{eqnarray}
where $A,A'$ denote dichotomic observables on the first qubit, $B,B'$ on the second qubit, and $C,C'$ on the third qubit. For local realistic models the bound $|\langle S \rangle| \leq 4$ holds, while quantum mechanics allows values up to $4\sqrt{2}$, certifying the presence of genuine tripartite nonlocality. 

In our construction, we select a specific realization of these measurement settings in terms of Pauli operators:
$A = \sigma_x,  A' = \sigma_y, 
B = \sigma_x,  B' = \sigma_y,
C = \tfrac{1}{\sqrt{2}}(\sigma_x + \sigma_y),  C' = \tfrac{1}{\sqrt{2}}(\sigma_x - \sigma_y).$
This choice generates eight distinct expectation values, $f_k = \langle \psi | S_k | \psi \rangle, \quad k=1,\ldots,8,$
where $S_k$ are the eight tensor-product terms appearing in (\ref{svetop}). These expectation values form the feature vector associated with each three-qubit state. The motivation behind this selection is that Svetlichny operators directly probe genuine multipartite entanglement and nonlocality, thereby providing features that are both physically interpretable and closely tied to the target variable of our learning model, namely the GME concurrence.

\subsection{Regression Models and Sampling Strategy}
In this work, we have used five state-of-the-art models (Support Vector Machine for Regression (SVM-R)\cite{vapnik1996support}, Decision Trees (DT-R)\cite{breiman2017classification}, Artificial Neural Network (ANN-R)\cite{specht1991general}, Generalized Additive Model (GAM)\cite{lou2012intelligible} and LS Boost-based Ensemble Model (LS-ENS)\cite{friedman2001greedy}) for regression. The models are applied to two datasets of quantum states generated from 2-qubit and 3-qubit states and estimate the concurrence of the quantum states. Also, each pipeline is assessed using 5-fold cross-validation sampling strategy. This subsection describes and contrasts the various classifiers used in the ML pipeline.\\
The selected models represent a diverse range of machine learning paradigms. The SVM-R is designed to achieve robust generalization by determining an optimal hyperplane that minimizes prediction error within an epsilon-insensitive margin, making it effective for non-linear regression problems. The DT-R, in contrast, constructs a hierarchical series of rules by recursively partitioning the input space, yielding simplicity, interpretability and transparency but showing a tendency toward overfitting in the absence of regularization. \\
ANN-R, with their layered and non-linear structure, serves as a universal function approximator capable of accurately modeling intricate patterns, though they demand extensive training data and parameter tuning. GAM extends regression through a probabilistic framework that adapts to complex data distributions, offering an advantage in capturing non-linear relationships among features, although at the cost of increased computational resources. Ensemble regression model with boosting approach, leverage the collective strength of multiple base learners to enhance prediction stability, reduce variance, and generally achieve superior performance compared to standalone models. The Ensemble-based methods strike a balance by combining multiple models to improve both accuracy and robustness, often mitigating overfitting while retaining predictive power. The misclassified observations are fitted in new learners and MSE is minimized between the observed response and the aggregated prediction of all previously grown learners.

\section{Experimental Setup}
The experiments were performed on a system with Windows 10 operating system, 16GB RAM and i7 (9th generation) Processor. MATLAB 2021 (64-bit) was used for conducting the experiment for the framework. The MATLAB toolbox for quantum entanglement, known as QETLAB, has been used for feature extraction and generation of target values (\cite{qetlab}).\\
The experiments were conducted using various regression models - SVM-R, DT-R, ANN-R, GAM, LS-ENS and five-fold cross validation sampling strategy. All model parameters were optimized using the hyperparameter tuning framework in MATLAB. We have used Root Mean Squared Error, Mean Absolute Error, Correlation coefficient and Coefficient of Determination as metrics to compare the performance of regression models. 
\begin{itemize}
\item Root Mean Squared Error (RMSE) measures the expected squared difference between actual ($y_i$)and predicted values($\hat{y}_{i}$), emphasizing larger errors. It is computed as $\text{RMSE} = \sqrt{\frac{1}{n} \sum_{i=1}^{n} (y_{i} - \hat{y}_{i})^{2}}$, where n is the total number of samples.
\item Mean Absolute Error (MAE) is a robust metric to measure the deviation of predicted values from actual ($y_i$) ones and is calculated as $\text{MAE} = \frac{1}{n} \sum_{i=1}^{n} |y_{i} - \hat{y}_{i})|$.

\item Correlation coefficient (R) indicates the strength and direction of the linear relationship between observed ($y_i$) and predicted values ($\hat{y}_i$). It is computed as $R = \frac{\sum_{i=1}^{n} (y_i - \bar{y}) (\hat{y}_i - \overline{\hat{y}})}{\sqrt{\sum_{i=1}^{n} (y_i - \bar{y})^2} \sqrt{\sum_{i=1}^{n} (\hat{y}_i - \overline{\hat{y}})^2}}$
\item Coefficient of Determination\(R^{2}\) represents the proportion of the variance in the dependent variable explained by the independent variable(s) in a regression model. It is computed as $R^2 = 1 - \sum_{i=1}^{n}(\hat{y}_i - \overline{\hat{y}})^2/\sum_{i=1}^{n}(y_i-\frac{1}{n}\sum_{i=1}^{n}y_i)^2$
\end{itemize}

\section{Results and Discussion}
\begin{figure}
    \centering
    \subfigure[DT-R]{\includegraphics[scale = 0.22]{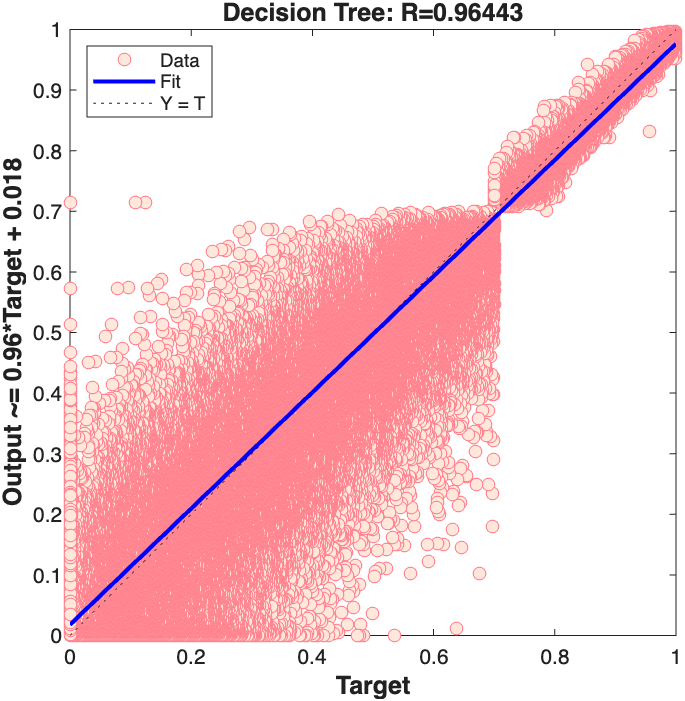}} 
    \subfigure[GAM]{\includegraphics[scale = 0.22]{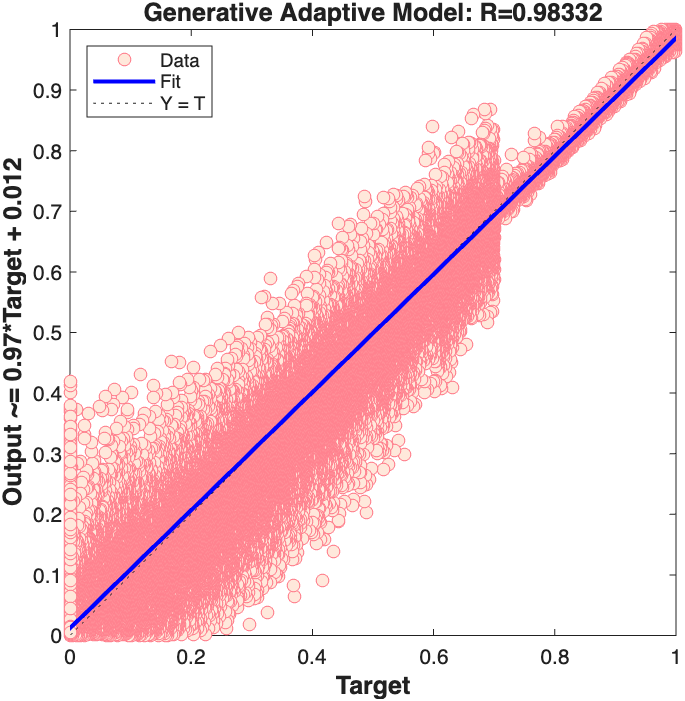}}
    \subfigure[SVM-R (Gaussian Kernel)]{\includegraphics[scale = 0.22]{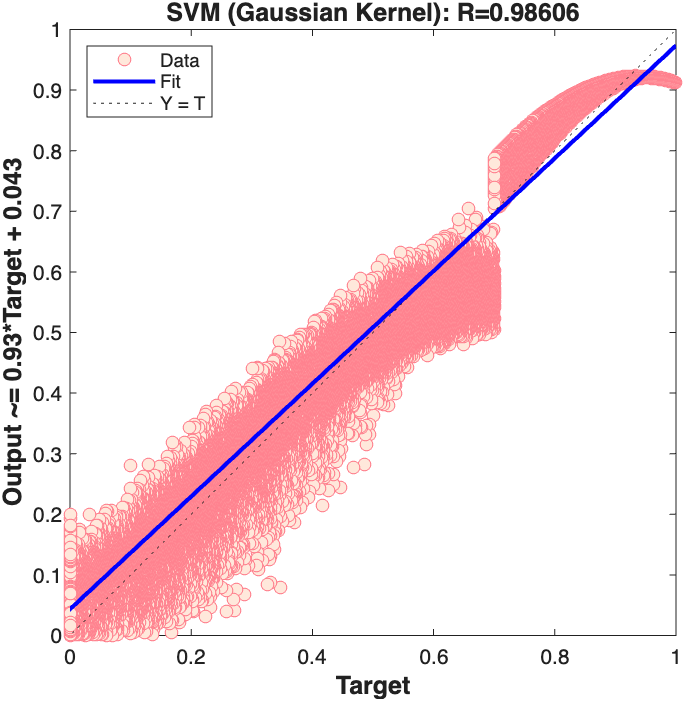}} 
    \subfigure[LS-ENS]{\includegraphics[scale = 0.22]{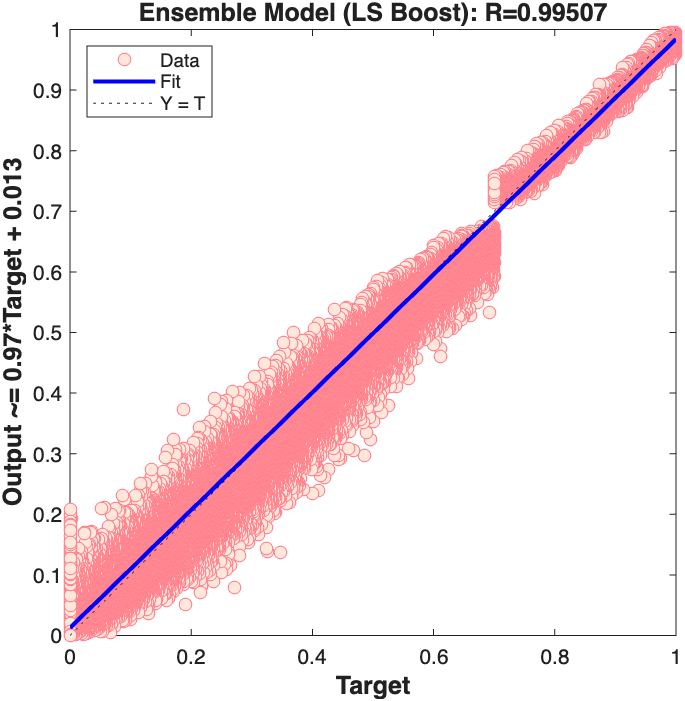}}
    \subfigure[ANN-R]{\includegraphics[scale = 0.22]{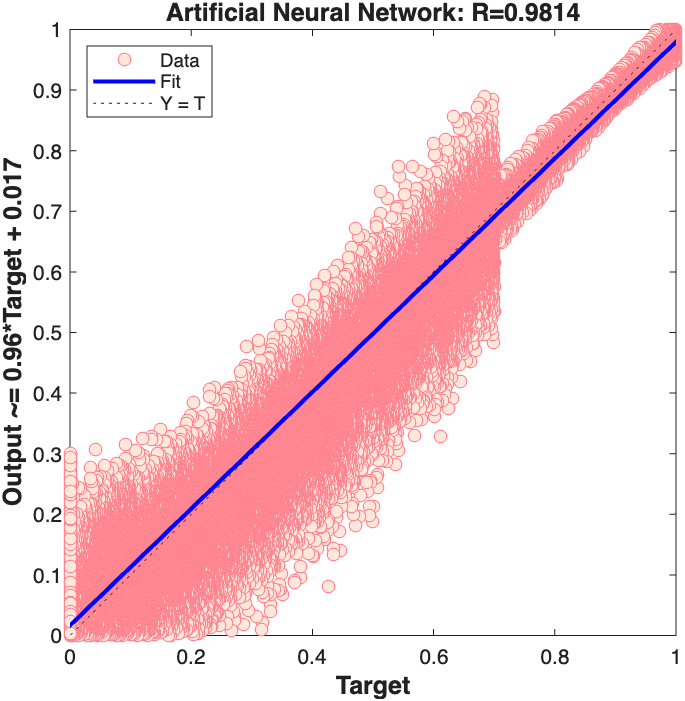}}
        \caption{Comparison between the analytically calculated concurrence and the concurrence predicted by different ML models for two-qubit system. Target value on $x$-axis represents the true concurrence, and the output on $y$-axis represents the predicted concurrence. The pink dots represent the sampled data, the dotted line represents the theoretical line where the predicted results are exactly equal to the true results, the sold blue line represents the fitting line, and R is the correlation coefficient.}
    \label{f1}
\end{figure}

\begin{figure}
    \centering
    \subfigure[DT-R]{\includegraphics[scale = 0.22]{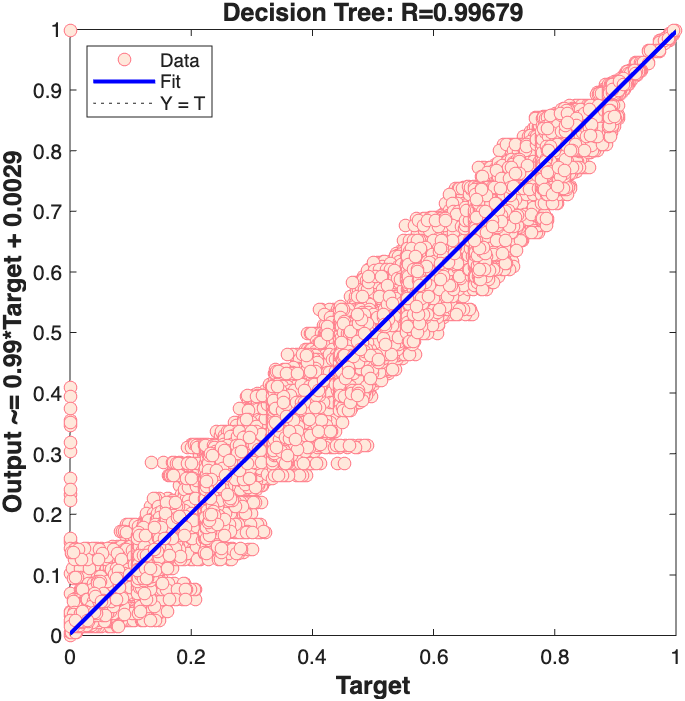}} 
    \subfigure[GAM]{\includegraphics[scale = 0.22]{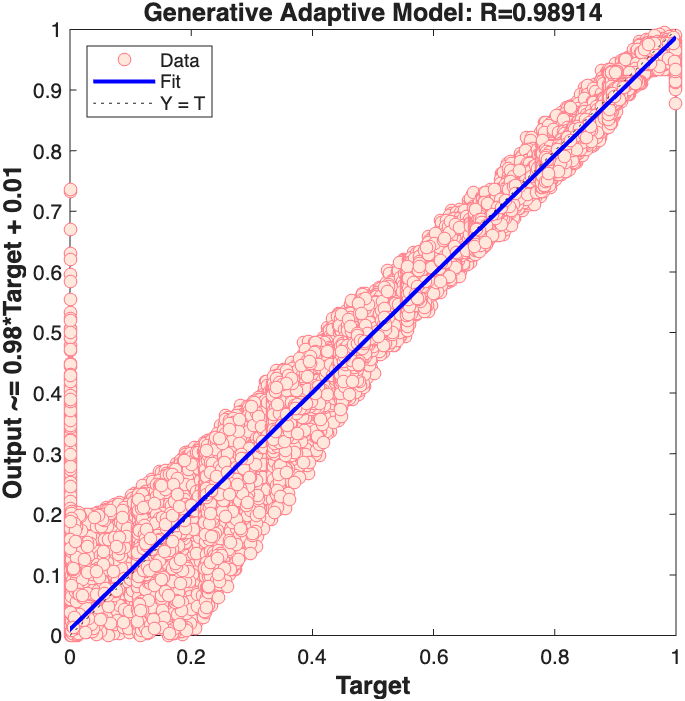}}
    \subfigure[SVM-R (Gaussian Kernel)]{\includegraphics[scale = 0.22]{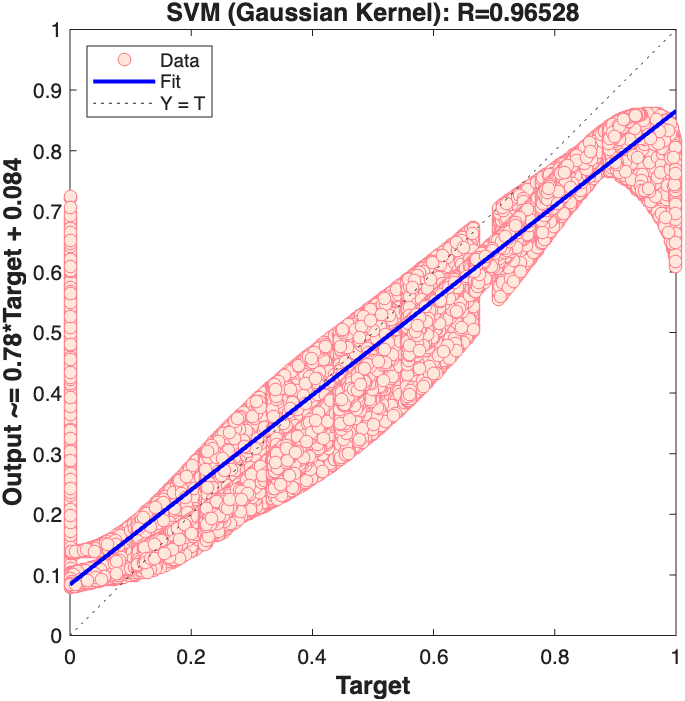}} 
    \subfigure[LS-ENS]{\includegraphics[scale = 0.22]{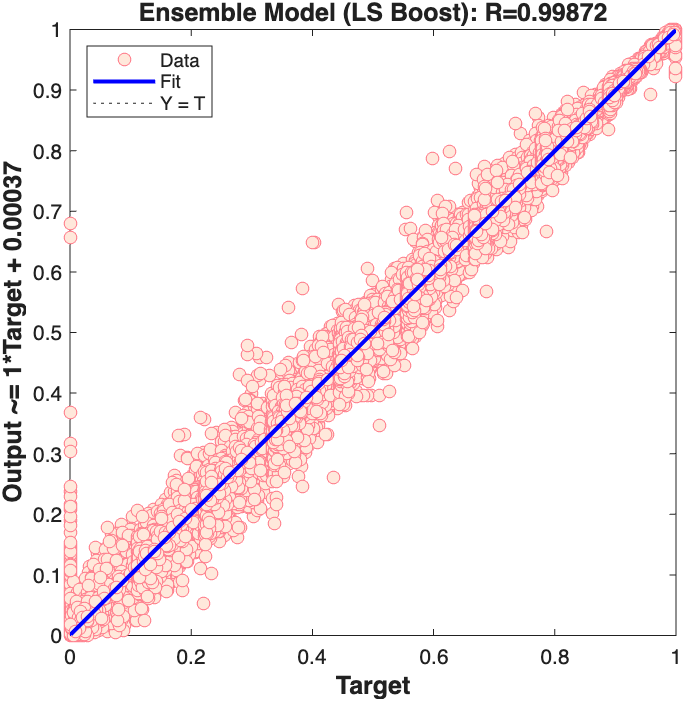}}
    \subfigure[ANN-R]{\includegraphics[scale = 0.22]{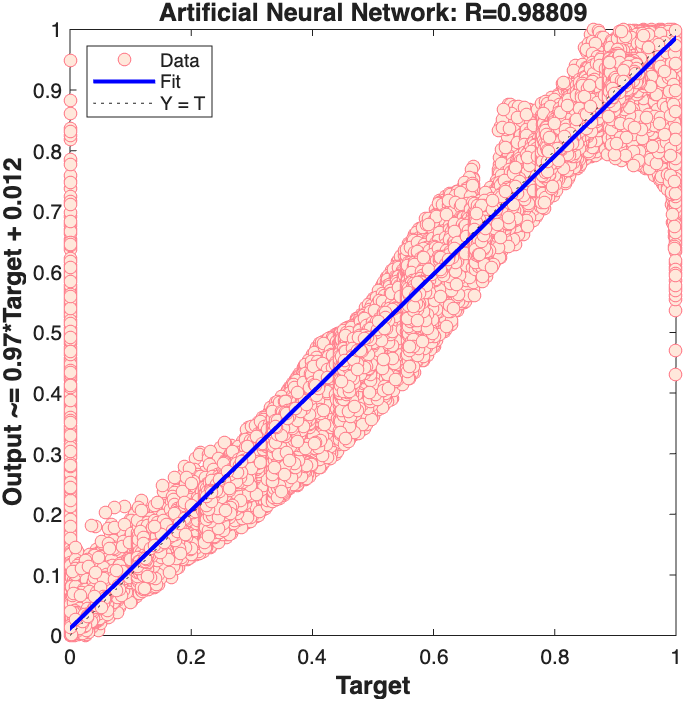}}
        \caption{Comparison between the analytically calculated GME concurrence and the concurrence predicted by different ML models for three-qubit system. Target value on $x$-axis represents the true GME concurrence, and the output on $y$-axis represents the predicted GME concurrence. The pink dots represent the sampled data, the dotted line represents the theoretical line where the predicted results are exactly equal to the true results, the sold blue line represents the fitting line, and R is the correlation coefficient.}
    \label{f2}
\end{figure}

This section states and discusses the performance of various regression models on predicting concurrence and GME concurrence for datasets of two and three-qubit states, respectively. The experiments conducted on both the datasets are independent and not related to each other in any respect. Table \ref{tab3} states the predictive performance for both datasets in terms of RMSE, MAE, R, and $R^2$. The Following observations were made from Table \ref{tab3}, Figures \ref{f1} and \ref{f2}:
\begin{table}[h]
\caption{Performance comparison of various Regression models for measuring Entanglement using concurrence and GME concurrence for datasets of two and three-qubit states }\label{tab3}
\begin{tabular}{@{}lccccp{0.5cm}cccc}
\toprule
&\multicolumn{4}{c}{\textbf{2-Qubit dataset}}&&\multicolumn{4}{c}{\textbf{3-Qubit dataset}}\\
\cmidrule{2-5}\cmidrule{7-10}
  \textbf{ Model Name} & \textbf{RMSE}& \textbf{MAE}&\textbf{R}& \textbf{$R^2$}&&\textbf{RMSE}&  \textbf{MAE}& \textbf{R}& \textbf{$R^2$}\\
\midrule
SVM-R (Gaussian kernel) & 0.0547        &  0.0443          &  0.9861        &  0.9691  &&  0.0960  &          0.0731  &     0.9653          &  0.9054\\
ANN-R                   & 0.0597        &  0.0426          &  0.9814        &  0.9631  &&  0.0480  &          0.0266  &     0.9881          &  0.9763\\
DT-R                    & 0.0828        &  0.0555          &  0.9644        &  0.9292  &&  0.0250  &          0.0135  &     0.9968          &  0.9936\\
LS-ENS                  & 0.0543  &\textbf{0.0429}         &  0.9951  &\textbf{0.9695} &&  0.0158  &  \textbf{0.0091} &     0.9987 &   \textbf{0.9974}\\
GAM                     & 0.0459        &  0.0258          &  0.9833        &  0.9784  &&  0.0408  &          0.0258  &     0.9891          &  0.9829\\
\botrule
\end{tabular}
\end{table}

\begin{itemize}
\item The performance of LS-ENS performed best among the regression models used for predicting the concurrence of 2\&3-qubit quantum states. Also, the performance of SVM-R was comparable to that of LS-ENS for 2-qubit quantum states.
\item The performance of DT-R was comparable among the regression models used for predicting the GME concurrence of 3-qubit quantum states and worst for predicting the Concurrence for 2-qubit quantum states.
\item The performance of majority of the models did not degrade with increase in the number of qubits. 
\item  Averaging the performance of regression models across the datasets, LS-ENS performed best.

\end{itemize}
    
\section{Conclusion and Future Scope}\label{sec13}
To summarize, we have presented an ML framework for quantifying entanglement in unknown quantum states. In this study, we have used five machine learning models: Decision Tree (DT), Generalized Additive Model (GAM), Support Vector Machines (SVM), LS Boost-based Ensemble Model (LS-ENS), Artificial Neural Networks (ANN). Using expectation values of measurements as input features, Concurrence and GME concurrence as target labels, these models were trained as entanglement quantifiers. Unlike conventional tomography based approaches, our method requires only partial measurement data, making it experimentally simpler and computationally faster. The results show that all five models generalize well relatively, demonstrating the versatility of machine learning for efficient entanglement quantification. \\
This work can be extended to higher-dimensional systems  where entanglement characterization is more complex. Exploring hybrid or unsupervised learning models with feature extraction and reduction methods and studying the impact of experimental noise on model accuracy would further improve their practical applicability. Moreover, integrating these models with real-time quantum device data could enable efficient, measurement-based entanglement estimation.

\bmhead{Acknowledgements} 
S.A. acknowledges financial support by Council of Scientific and Industrial Research, Government of India (Grant No. 08/0133(23695)/2025-EMR-I).

\bmhead{Declaration of Competing Interests}  
The authors declare that they have no known competing financial interests or personal relationships
that could have appeared to influence the work reported in this paper.

 \bmhead{Code availability} The code that supports the findings of this study is available upon reasonable request.

\bibliography{sn-bibliography}


\begin{thebibliography}{30}
\ifx \bisbn   \undefined \def \bisbn  #1{ISBN #1}\fi
\ifx \binits  \undefined \def \binits#1{#1}\fi
\ifx \bauthor  \undefined \def \bauthor#1{#1}\fi
\ifx \batitle  \undefined \def \batitle#1{#1}\fi
\ifx \bjtitle  \undefined \def \bjtitle#1{#1}\fi
\ifx \bvolume  \undefined \def \bvolume#1{\textbf{#1}}\fi
\ifx \byear  \undefined \def \byear#1{#1}\fi
\ifx \bissue  \undefined \def \bissue#1{#1}\fi
\ifx \bfpage  \undefined \def \bfpage#1{#1}\fi
\ifx \blpage  \undefined \def \blpage #1{#1}\fi
\ifx \burl  \undefined \def \burl#1{\textsf{#1}}\fi
\ifx \doiurl  \undefined \def \doiurl#1{\url{https://doi.org/#1}}\fi
\ifx \betal  \undefined \def \betal{\textit{et al.}}\fi
\ifx \binstitute  \undefined \def \binstitute#1{#1}\fi
\ifx \binstitutionaled  \undefined \def \binstitutionaled#1{#1}\fi
\ifx \bctitle  \undefined \def \bctitle#1{#1}\fi
\ifx \beditor  \undefined \def \beditor#1{#1}\fi
\ifx \bpublisher  \undefined \def \bpublisher#1{#1}\fi
\ifx \bbtitle  \undefined \def \bbtitle#1{#1}\fi
\ifx \bedition  \undefined \def \bedition#1{#1}\fi
\ifx \bseriesno  \undefined \def \bseriesno#1{#1}\fi
\ifx \blocation  \undefined \def \blocation#1{#1}\fi
\ifx \bsertitle  \undefined \def \bsertitle#1{#1}\fi
\ifx \bsnm \undefined \def \bsnm#1{#1}\fi
\ifx \bsuffix \undefined \def \bsuffix#1{#1}\fi
\ifx \bparticle \undefined \def \bparticle#1{#1}\fi
\ifx \barticle \undefined \def \barticle#1{#1}\fi
\bibcommenthead
\ifx \bconfdate \undefined \def \bconfdate #1{#1}\fi
\ifx \botherref \undefined \def \botherref #1{#1}\fi
\ifx \url \undefined \def \url#1{\textsf{#1}}\fi
\ifx \bchapter \undefined \def \bchapter#1{#1}\fi
\ifx \bbook \undefined \def \bbook#1{#1}\fi
\ifx \bcomment \undefined \def \bcomment#1{#1}\fi
\ifx \oauthor \undefined \def \oauthor#1{#1}\fi
\ifx \citeauthoryear \undefined \def \citeauthoryear#1{#1}\fi
\ifx \endbibitem  \undefined \def \endbibitem {}\fi
\ifx \bconflocation  \undefined \def \bconflocation#1{#1}\fi
\ifx \arxivurl  \undefined \def \arxivurl#1{\textsf{#1}}\fi
\csname PreBibitemsHook\endcsname

\bibitem[\protect\citeauthoryear{Bennett and
  Wiesner}{1992}]{PhysRevLett.69.2881}
\begin{barticle}
\bauthor{\bsnm{Bennett}, \binits{C.H.}},
\bauthor{\bsnm{Wiesner}, \binits{S.J.}}:
\batitle{Communication via one- and two-particle operators on
  einstein-podolsky-rosen states}.
\bjtitle{Phys. Rev. Lett.}
\bvolume{69},
\bfpage{2881}--\blpage{2884}
(\byear{1992})
\doiurl{10.1103/PhysRevLett.69.2881}
\end{barticle}
\endbibitem

\bibitem[\protect\citeauthoryear{Bennett et~al.}{1993}]{PhysRevLett.70.1895}
\begin{barticle}
\bauthor{\bsnm{Bennett}, \binits{C.H.}},
\bauthor{\bsnm{Brassard}, \binits{G.}},
\bauthor{\bsnm{Cr\'epeau}, \binits{C.}},
\bauthor{\bsnm{Jozsa}, \binits{R.}},
\bauthor{\bsnm{Peres}, \binits{A.}},
\bauthor{\bsnm{Wootters}, \binits{W.K.}}:
\batitle{Teleporting an unknown quantum state via dual classical and
  einstein-podolsky-rosen channels}.
\bjtitle{Phys. Rev. Lett.}
\bvolume{70},
\bfpage{1895}--\blpage{1899}
(\byear{1993})
\doiurl{10.1103/PhysRevLett.70.1895}
\end{barticle}
\endbibitem

\bibitem[\protect\citeauthoryear{Ekert}{1991}]{PhysRevLett.67.661}
\begin{barticle}
\bauthor{\bsnm{Ekert}, \binits{A.K.}}:
\batitle{Quantum cryptography based on bell's theorem}.
\bjtitle{Phys. Rev. Lett.}
\bvolume{67},
\bfpage{661}--\blpage{663}
(\byear{1991})
\doiurl{10.1103/PhysRevLett.67.661}
\end{barticle}
\endbibitem

\bibitem[\protect\citeauthoryear{Horodecki et~al.}{2009}]{RevModPhys.81.865}
\begin{barticle}
\bauthor{\bsnm{Horodecki}, \binits{R.}},
\bauthor{\bsnm{Horodecki}, \binits{P.}},
\bauthor{\bsnm{Horodecki}, \binits{M.}},
\bauthor{\bsnm{Horodecki}, \binits{K.}}:
\batitle{Quantum entanglement}.
\bjtitle{Rev. Mod. Phys.}
\bvolume{81},
\bfpage{865}--\blpage{942}
(\byear{2009})
\doiurl{10.1103/RevModPhys.81.865}
\end{barticle}
\endbibitem

\bibitem[\protect\citeauthoryear{Ma et~al.}{2011}]{Ma2011_GME}
\begin{barticle}
\bauthor{\bsnm{Ma}, \binits{Z.-H.}},
\bauthor{\bsnm{Chen}, \binits{Z.-H.}},
\bauthor{\bsnm{Chen}, \binits{J.-L.}},
\bauthor{\bsnm{Spengler}, \binits{C.}},
\bauthor{\bsnm{Gabriel}, \binits{A.}},
\bauthor{\bsnm{Huber}, \binits{M.}}:
\batitle{Measure of genuine multipartite entanglement with computable lower
  bounds}.
\bjtitle{Physical Review A}
\bvolume{83}(\bissue{6}),
\bfpage{062325}
(\byear{2011})
\doiurl{10.1103/PhysRevA.83.062325}
\end{barticle}
\endbibitem

\bibitem[\protect\citeauthoryear{Aggarwal}{2025}]{aggarwal2025classification}
\begin{botherref}
\oauthor{\bsnm{Aggarwal}, \binits{S.}}:
Classification of three-qubit genuine entangled states using concurrence fill.
arXiv preprint arXiv:2506.16935
(2025)
\end{botherref}
\endbibitem

\bibitem[\protect\citeauthoryear{Wootters}{1998}]{woottersprl98}
\begin{barticle}
\bauthor{\bsnm{Wootters}, \binits{W.K.}}:
\batitle{Entanglement of formation of an arbitrary state of two qubits}.
\bjtitle{Phys. Rev. Lett.}
\bvolume{80},
\bfpage{2245}--\blpage{2248}
(\byear{1998})
\doiurl{10.1103/PhysRevLett.80.2245}
\end{barticle}
\endbibitem

\bibitem[\protect\citeauthoryear{Hill and Wootters}{1997}]{hillprl97}
\begin{barticle}
\bauthor{\bsnm{Hill}, \binits{S.A.}},
\bauthor{\bsnm{Wootters}, \binits{W.K.}}:
\batitle{Entanglement of a pair of quantum bits}.
\bjtitle{Phys. Rev. Lett.}
\bvolume{78},
\bfpage{5022}--\blpage{5025}
(\byear{1997})
\doiurl{10.1103/PhysRevLett.78.5022}
\end{barticle}
\endbibitem

\bibitem[\protect\citeauthoryear{Vidal and Werner}{2002}]{vidalpra2002}
\begin{barticle}
\bauthor{\bsnm{Vidal}, \binits{G.}},
\bauthor{\bsnm{Werner}, \binits{R.F.}}:
\batitle{Computable measure of entanglement}.
\bjtitle{Phys. Rev. A}
\bvolume{65},
\bfpage{032314}
(\byear{2002})
\doiurl{10.1103/PhysRevA.65.032314}
\end{barticle}
\endbibitem

\bibitem[\protect\citeauthoryear{Kim}{2010}]{kimpra2010}
\begin{barticle}
\bauthor{\bsnm{Kim}, \binits{J.S.}}:
\batitle{Tsallis entropy and entanglement constraints in multiqubit systems}.
\bjtitle{Phys. Rev. A}
\bvolume{81},
\bfpage{062328}
(\byear{2010})
\doiurl{10.1103/PhysRevA.81.062328}
\end{barticle}
\endbibitem

\bibitem[\protect\citeauthoryear{Krammer et~al.}{2009}]{krammerprl2009}
\begin{barticle}
\bauthor{\bsnm{Krammer}, \binits{P.}},
\bauthor{\bsnm{Kampermann}, \binits{H.}},
\bauthor{\bsnm{Bru\ss{}}, \binits{D.}},
\bauthor{\bsnm{Bertlmann}, \binits{R.A.}},
\bauthor{\bsnm{Kwek}, \binits{L.C.}},
\bauthor{\bsnm{Macchiavello}, \binits{C.}}:
\batitle{Multipartite entanglement detection via structure factors}.
\bjtitle{Phys. Rev. Lett.}
\bvolume{103},
\bfpage{100502}
(\byear{2009})
\doiurl{10.1103/PhysRevLett.103.100502}
\end{barticle}
\endbibitem

\bibitem[\protect\citeauthoryear{Schmidt}{1907}]{schmidt1907theorie}
\begin{barticle}
\bauthor{\bsnm{Schmidt}, \binits{E.}}:
\batitle{Zur theorie der linearen und nichtlinearen integralgleichungen}.
\bjtitle{Mathematische Annalen}
\bvolume{63}(\bissue{4}),
\bfpage{433}--\blpage{476}
(\byear{1907})
\end{barticle}
\endbibitem

\bibitem[\protect\citeauthoryear{Bennett et~al.}{1996}]{bennetprl96}
\begin{barticle}
\bauthor{\bsnm{Bennett}, \binits{C.H.}},
\bauthor{\bsnm{Bernstein}, \binits{H.J.}},
\bauthor{\bsnm{Popescu}, \binits{S.}},
\bauthor{\bsnm{Schumacher}, \binits{B.}}:
\batitle{Concentrating partial entanglement by local operations}.
\bjtitle{Phys. Rev. A}
\bvolume{53},
\bfpage{2046}--\blpage{2052}
(\byear{1996})
\doiurl{10.1103/PhysRevA.53.2046}
\end{barticle}
\endbibitem

\bibitem[\protect\citeauthoryear{Ac\'{\i}n et~al.}{2000}]{tarrachprl2000}
\begin{barticle}
\bauthor{\bsnm{Ac\'{\i}n}, \binits{A.}},
\bauthor{\bsnm{Andrianov}, \binits{A.}},
\bauthor{\bsnm{Costa}, \binits{L.}},
\bauthor{\bsnm{Jan\'e}, \binits{E.}},
\bauthor{\bsnm{Latorre}, \binits{J.I.}},
\bauthor{\bsnm{Tarrach}, \binits{R.}}:
\batitle{Generalized schmidt decomposition and classification of
  three-quantum-bit states}.
\bjtitle{Phys. Rev. Lett.}
\bvolume{85},
\bfpage{1560}--\blpage{1563}
(\byear{2000})
\doiurl{10.1103/PhysRevLett.85.1560}
\end{barticle}
\endbibitem

\bibitem[\protect\citeauthoryear{Koutn{\'y} et~al.}{2023}]{Koutny2023}
\begin{barticle}
\bauthor{\bsnm{Koutn{\'y}}, \binits{D.}},
\bauthor{\bsnm{B{\"a}rtschi}, \binits{A.}},
\bauthor{\bsnm{Flammia}, \binits{S.T.}},
\bauthor{\bsnm{Bravyi}, \binits{S.}},
\bauthor{\bsnm{Gambetta}, \binits{J.M.}},
\bauthor{\bsnm{Temme}, \binits{K.}}:
\batitle{Detecting and quantifying entanglement with machine learning}.
\bjtitle{Nature Communications}
\bvolume{14},
\bfpage{3801}
(\byear{2023})
\doiurl{10.1038/s41467-023-39542-2}
\end{barticle}
\endbibitem

\bibitem[\protect\citeauthoryear{Wang et~al.}{2025}]{Wang2025}
\begin{barticle}
\bauthor{\bsnm{Wang}, \binits{X.}},
\bauthor{\bsnm{Li}, \binits{Y.}},
\bauthor{\bsnm{Zhang}, \binits{H.}}:
\batitle{Quantifying multipartite entanglement with machine learning
  approaches}.
\bjtitle{Entropy}
\bvolume{27}(\bissue{2}),
\bfpage{185}
(\byear{2025})
\doiurl{10.3390/e27020185}
\end{barticle}
\endbibitem

\bibitem[\protect\citeauthoryear{Feng and Chen}{2024}]{feng2024quantifying}
\begin{barticle}
\bauthor{\bsnm{Feng}, \binits{C.}},
\bauthor{\bsnm{Chen}, \binits{L.}}:
\batitle{Quantifying quantum entanglement via machine learning models}.
\bjtitle{Communications in Theoretical Physics}
\bvolume{76}(\bissue{7}),
\bfpage{075104}
(\byear{2024})
\end{barticle}
\endbibitem

\bibitem[\protect\citeauthoryear{Mahdian and
  Mousavi}{2025}]{mahdian2025entanglement}
\begin{barticle}
\bauthor{\bsnm{Mahdian}, \binits{M.}},
\bauthor{\bsnm{Mousavi}, \binits{Z.}}:
\batitle{Entanglement detection with quantum support vector machine (qsvm) on
  near-term quantum devices}.
\bjtitle{Scientific Reports}
\bvolume{15}(\bissue{1}),
\bfpage{1}--\blpage{15}
(\byear{2025})
\end{barticle}
\endbibitem

\bibitem[\protect\citeauthoryear{Huang et~al.}{2025}]{huang2025direct}
\begin{barticle}
\bauthor{\bsnm{Huang}, \binits{Y.}},
\bauthor{\bsnm{Che}, \binits{L.}},
\bauthor{\bsnm{Wei}, \binits{C.}},
\bauthor{\bsnm{Xu}, \binits{F.}},
\bauthor{\bsnm{Nie}, \binits{X.}},
\bauthor{\bsnm{Li}, \binits{J.}},
\bauthor{\bsnm{Lu}, \binits{D.}},
\bauthor{\bsnm{Xin}, \binits{T.}}:
\batitle{Direct entanglement detection of quantum systems using machine
  learning}.
\bjtitle{npj Quantum Information}
\bvolume{11}(\bissue{1}),
\bfpage{29}
(\byear{2025})
\end{barticle}
\endbibitem

\bibitem[\protect\citeauthoryear{Vapnik et~al.}{1996}]{vapnik1996support}
\begin{botherref}
\oauthor{\bsnm{Vapnik}, \binits{V.}},
\oauthor{\bsnm{Golowich}, \binits{S.}},
\oauthor{\bsnm{Smola}, \binits{A.}}:
Support vector method for function approximation, regression estimation and
  signal processing.
Advances in neural information processing systems
\textbf{9}
(1996)
\end{botherref}
\endbibitem

\bibitem[\protect\citeauthoryear{Breiman
  et~al.}{2017}]{breiman2017classification}
\begin{bbook}
\bauthor{\bsnm{Breiman}, \binits{L.}},
\bauthor{\bsnm{Friedman}, \binits{J.}},
\bauthor{\bsnm{Olshen}, \binits{R.A.}},
\bauthor{\bsnm{Stone}, \binits{C.J.}}:
\bbtitle{Classification and Regression Trees}.
\bpublisher{Chapman and Hall/CRC}, \blocation{???}
(\byear{2017})
\end{bbook}
\endbibitem

\bibitem[\protect\citeauthoryear{Specht et~al.}{1991}]{specht1991general}
\begin{barticle}
\bauthor{\bsnm{Specht}, \binits{D.F.}}, \betal:
\batitle{A general regression neural network}.
\bjtitle{IEEE transactions on neural networks}
\bvolume{2}(\bissue{6}),
\bfpage{568}--\blpage{576}
(\byear{1991})
\end{barticle}
\endbibitem

\bibitem[\protect\citeauthoryear{Friedman}{2001}]{friedman2001greedy}
\begin{botherref}
\oauthor{\bsnm{Friedman}, \binits{J.H.}}:
Greedy function approximation: a gradient boosting machine.
Annals of statistics,
1189--1232
(2001)
\end{botherref}
\endbibitem

\bibitem[\protect\citeauthoryear{Ac{\'i}n et~al.}{2001}]{acin2001}
\begin{barticle}
\bauthor{\bsnm{Ac{\'i}n}, \binits{A.}},
\bauthor{\bsnm{Bru{\ss}}, \binits{D.}},
\bauthor{\bsnm{Lewenstein}, \binits{M.}},
\bauthor{\bsnm{Sanpera}, \binits{A.}}:
\batitle{Classification of mixed three-qubit states}.
\bjtitle{Physical Review Letters}
\bvolume{87}(\bissue{4}),
\bfpage{040401}
(\byear{2001})
\doiurl{10.1103/PhysRevLett.87.040401}
\end{barticle}
\endbibitem

\bibitem[\protect\citeauthoryear{Aggarwal and
  Adhikari}{2021}]{aggarwal2021witness}
\begin{botherref}
\oauthor{\bsnm{Aggarwal}, \binits{S.}},
\oauthor{\bsnm{Adhikari}, \binits{S.}}:
Witness operator provides better estimate of the lower bound of concurrence of
  bipartite bound entangled states in $d_1 \otimes d_2$ dimensional system.
Quantum Information Processing
\textbf{20}(3)
(2021)
\doiurl{10.1007/s11128-021-03012-4}
\end{botherref}
\endbibitem

\bibitem[\protect\citeauthoryear{Pan et~al.}{2024}]{pan2024quantifying}
\begin{barticle}
\bauthor{\bsnm{Pan}, \binits{G.-Z.}},
\bauthor{\bsnm{Yang}, \binits{M.}},
\bauthor{\bsnm{Zhou}, \binits{J.}},
\bauthor{\bsnm{Yuan}, \binits{H.}},
\bauthor{\bsnm{Miao}, \binits{C.}},
\bauthor{\bsnm{Zhang}, \binits{G.}}:
\batitle{Quantifying entanglement for unknown quantum states via artificial
  neural networks}.
\bjtitle{Scientific Reports}
\bvolume{14}(\bissue{1}),
\bfpage{26267}
(\byear{2024})
\end{barticle}
\endbibitem

\bibitem[\protect\citeauthoryear{Lin et~al.}{2023}]{lin2023quantifying}
\begin{barticle}
\bauthor{\bsnm{Lin}, \binits{X.}},
\bauthor{\bsnm{Chen}, \binits{Z.}},
\bauthor{\bsnm{Wei}, \binits{Z.}}:
\batitle{Quantifying quantum entanglement via a hybrid quantum-classical
  machine learning framework}.
\bjtitle{Physical Review A}
\bvolume{107}(\bissue{6}),
\bfpage{062409}
(\byear{2023})
\end{barticle}
\endbibitem

\bibitem[\protect\citeauthoryear{Svetlichny}{1987}]{PhysRevD.35.3066}
\begin{barticle}
\bauthor{\bsnm{Svetlichny}, \binits{G.}}:
\batitle{Distinguishing three-body from two-body nonseparability by a bell-type
  inequality}.
\bjtitle{Phys. Rev. D}
\bvolume{35},
\bfpage{3066}--\blpage{3069}
(\byear{1987})
\doiurl{10.1103/PhysRevD.35.3066}
\end{barticle}
\endbibitem

\bibitem[\protect\citeauthoryear{Lou et~al.}{2012}]{lou2012intelligible}
\begin{bchapter}
\bauthor{\bsnm{Lou}, \binits{Y.}},
\bauthor{\bsnm{Caruana}, \binits{R.}},
\bauthor{\bsnm{Gehrke}, \binits{J.}}:
\bctitle{Intelligible models for classification and regression}.
In: \bbtitle{Proceedings of the 18th ACM SIGKDD International Conference on
  Knowledge Discovery and Data Mining},
pp. \bfpage{150}--\blpage{158}
(\byear{2012})
\end{bchapter}
\endbibitem

\bibitem[\protect\citeauthoryear{Johnston}{2016}]{qetlab}
\begin{botherref}
\oauthor{\bsnm{Johnston}, \binits{N.}}:
{QETLAB}: A {MATLAB} toolbox for quantum entanglement, version 1.0.
\url{https://qetlab.com}
(2016).
\doiurl{10.5281/zenodo.44637}
\end{botherref}
\endbibitem

\end{thebibliography}
\end{document}